\documentclass[final]{elsarticle}

\usepackage{color}

\usepackage{graphics}
\usepackage{subfigure}

\usepackage{amssymb}

\journal{Applied Mathematics and Computation}

\begin{document}

\begin{frontmatter}



\title{A Local Discontinuous Galerkin Method for 1.5-Dimensional Streamer Discharge Simulations}


\author{Chijie Zhuang, Rong Zeng}
\address{Department of Electrical Engineering, Tsinghua University, Beijing 100084, China}

\begin{abstract}
Streamer discharges are important both in theory and industry applications. This paper proposed a local discontinuous Galerkin
method to simulate the convection dominated fluid model of streamer discharges. To simulate the rapid transient streamer discharge
process, a method with high resolution and high order accuracy is highly desired. Combining the advantages of finite volume and
finite element method, local discontinuous Galerkin method is such a choice. In this paper, a simulation of a double-headed streamer discharge in
nitrogen was performed {\color{black}by} using 1.5-dimensional fluid model. The preliminary results indicate the potential of extending the method to
general streamer simulations in complex geometries.
\end{abstract}


\end{frontmatter}


Streamers occur in nature as well as in many industrial applications.
They are a generic initial stage of sparks, lighting, and various
other discharges. The available experimental
results and measuring methods for streamers discharge research are still insufficient to
build a detailed picture of streamer development,
which makes numerical simulations essential tools for a better understanding of streamer physics.

The simplest and most frequently used model for streamer discharge
is the fluid model, which consists of two continuity
equations (which are convection-dominated diffusion equations with source terms) coupled
with a Poisson's equation, see Eq. (\ref{eq1.1}) to Eq. (\ref{eq1.5}),
\begin{eqnarray}
& \frac{\partial n_e}{\partial t} + \nabla \cdot ({\vec v_{e}}n_e)-\nabla \cdot (D \nabla n_e)=\alpha n_e |\vec v_e|, \label{eq1.1}\\
& \frac{\partial n_p}{\partial t} + \nabla \cdot ({\vec v_{p}}n_p)= \alpha n_e |\vec v_e|, \label{eq1.2}\\
& \nabla \cdot (\varepsilon_0 \nabla U)= e_0 ( n_e - n_p),\\
& \vec E = (E_r, E_z)^T = -(\frac{\partial U}{\partial r}, \frac{\partial U}{\partial z})^T,~|\vec E|=\sqrt{E_r^2+E_z^2},\\
& \vec v_{e,p} = (v_{(e,p),r},v_{(e,p),z})^T= \mu_{e,p}\vec E,~|\vec v_e|=\sqrt{v_{e,r}^2+v_{e,z}^2}. \label{eq1.5}
\end{eqnarray}
where $n_{e,p}$ are the charged particle densities, $\mu_{e,p}$ are the movability coefficients, $\vec v_{e,p}$ are the convection velocities, $D$ is the diffusion coefficient, the index $e$, $p$ means electrons and positive ions, respectively. $U$ and $\vec E$ are the electrical potential and electric field, respectively; $\varepsilon_0$ is the dielectric coefficient in air; $e_0$ is the unit charge of an electron. $\alpha$ and $\eta$ are measured by experiments. See \cite{kunhart} for the {\color{black} details of the parameters}.

The continuity equations, i.e., Eq. (\ref{eq1.1}) and Eq. (\ref{eq1.2}), are convection dominated.
Traditional linear numerical schemes often
generate too much numerical diffusions or oscillations, which will be shown by a simple example in the Appendix.
A scheme free of numerical oscillations and of high resolution is greatly desired.
In addition, a high order scheme would be computationally more efficient than a low order one.

Several methods have been used for streamer discharge simulations since 1960s \cite{davies1}.
The two-step
Lax-Wendroff method was used to solve the continuity equations in 1981 \cite{morrow2},
but suffered from numerical
oscillations for high density plasmas. This drawback was overcome by
{\color{black}flux-corrected transport} (FCT)
technique \cite{zalezak}. Finite difference method (FDM) combined with FCT was widely used
during 1980s and 1990s for streamer discharge simulations \cite{kunhart}. But the finite difference characteristics made it hard to
handle unstructured grids.
Starting from 1990s, FEM-FCT \cite{lohner} was introduced to the field of streamer simulations \cite{georghiou}\cite{min2}. FEM can handle complex geometries and unstructured grids, which
dramatically reduces the degrees of freedom while maintains a comparable
accuracy as FD-FCT, resulting in large computational savings.
However, FEM does not enforce
the local conservation, which makes the total current law violated in streamer discharge simulations. To enforce local conservation, finite volume
method (FVM) may be a more natural choice and it becomes popular for streamer simulations since 2000 \cite{ebert1}\cite{moving mesh}. FVM can handle complex geometries, but it needs wide stencils to construct high order schemes. In addition, it is much easier to use FEM to discretize the diffusion term than FVM.

The brief literature reviewed above shows the desired properties of an ideal streamer discharge simulation method: being able to suppress non-physical oscillations, this is why FCT is introduced; flexibility with unstructured grids, this is why FEM is introduced; being able to enforce local conservation, this is why FVM is introduced; in addition, high order accuracy is a benefit.

Discontinuous Galerkin (DG) method seems to be such a
choice \cite{shureview}. It uses a finite element space discretization by
discontinuous approximations, and incorporates the ideas of
numerical fluxes and slope limiters from the high-resolution FDM and
FVM. The resulting method is compact, local conservative and high order accurate. In addition, it is easy to handle the diffusion term within the framework of discontinuous Galerkin method, e.g., the local discontinuous Galerkin method \cite{shureview}. In this paper, we show DG method a competitor with existing methods for streamer discharge simulations.

\section{The reduced 1.5-dimensional model}
Eq. (\ref{eq1}) to Eq. (\ref{eq5}) describe a 3-dimensional model. If we consider the axial symmetry of the model, i.e., $\frac{\partial n_e}{\partial \theta}=\frac{\partial n_p}{\partial \theta}=0$, the model can be reduced to a 2-dimensional model in cylindrical coordinate system. If we further assume that the charge is distributed in discs with a fixed radius $r_d$, and at each disc, the charge density is uniform, i.e., $\frac{\partial n_e}{\partial r}=\frac{\partial n_p}{\partial r}=0$, and the charges only move along the $z$-axis, the 2-dimensional model can be further reduced to a 1.5-dimensional model, i.e., solving Poisson's equation in 2-dimensional cylindrical coordinate system and the continuity equations in 1-dimension \cite{davies1} (cf. Figure \ref{fig:disc}). This is the model used by Davies \cite{davies1}. The 1.5-dimensional model is as follows:
{
\color{black}
\begin{eqnarray}
\frac{\partial n_e}{\partial t}+\frac{\partial (n_ev_{e,z})}{\partial z}-D\frac{\partial^2 n_e}{\partial z}=\alpha n_e|v_{e,z}|, \label{eq1} \\
\frac{\partial n_p}{\partial t}+\frac{\partial (n_pv_{p,z})}{\partial z} =\alpha n_e |v_{e,z}|,  \label{eqpc}\\
\Delta u = -\frac{e}{\varepsilon_0}(n_p-n_e),~~E_z=-\frac{\partial u}{\partial z}. \\
v_{e,z}= \mu_{e}E_z,~~ v_{p,z}=\mu_pE_z \label{eq5}
\end{eqnarray}
}

\section{Algorithm for the 1.5-dimensional model}
\subsection{Solution to Poisson's equation}
For the 1.5-dimensional model, the Possion's equation is solved analytically  by the disc method~\cite{davies1}. The electric field at the $z$-axis ($E_z$) equals to the sum of the field generated by the space charges ($E_s$), and the Laplace electric field ($E_{laplace}$):
\begin{equation}
E_z(z)=E_s(z)+E_{laplace}(z).
\end{equation}
$E_{laplace}$ is determined by the applied voltage.

\begin{figure}[!h]
\centering
\includegraphics[width=0.485\textwidth]{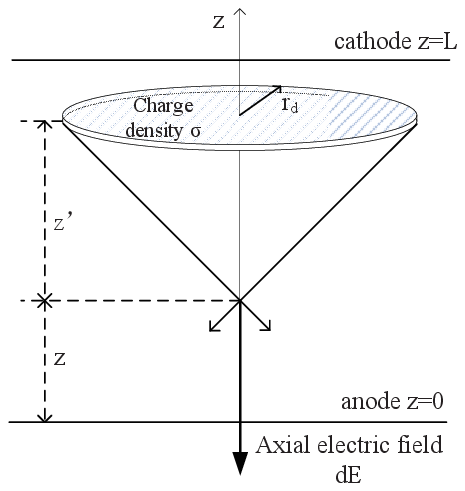} \\
  \caption{disc model and the configuration.}
  \label{fig:disc}
\end{figure}

Suppose that there is a disc with net charge
density $\sigma$, radius $r_d$, thickness $\mbox{d}z$, as shown in Figure \ref{fig:disc}, the electric field it generates at point $z$ along the
$z$-axis is:\\
{\color{black}
\begin{equation}
\mbox{d}E_s(z)=\Bigg\{
\begin{array}{l l}
\frac{1}{2\varepsilon_0}\sigma(z+z')(\frac{z'}{\sqrt{z'^2+r_d^2}}+1)\mbox{d}z,& z' < 0;\\
\frac{1}{2\varepsilon_0}\sigma(z+z')(
\frac{z'}{\sqrt{z'^2+r_d^2}}-1)\mbox{d}z,& z'\ge 0.
\end{array}
\end{equation}
}
where $z'$ means the {\color{black} signed} distance between the point $z$ and the {\color{black}center of the disc}.

To consider the influence of the electrodes to the electric field, the image charges should be taken into
account. We consider the image charges whose distance to the electrodes are less than $L$, where $L$ is the length of the discharge gap. Integrating over the whole domain, the solution of $E_s$ reads
\begin{eqnarray}
E_s(z)&=&\frac{1}{2\varepsilon_0}\bigg[\int_{-L}^{0} \sigma
(z+z')\bigg(\frac{z'}{\sqrt{z'^2+r_d^2}}+1\bigg)\mbox{d}z' \nonumber \\
&&+\int_{0}^{L}\sigma(z+z')\bigg(\frac{z'}{\sqrt{z'^2+r_d^2}}-1\bigg)\mbox{d}z'\bigg].
\end{eqnarray}

\subsection{Spacial discretization of the continuity equation}
For clarity, we only take Eq. (\ref{eq1}) as an example to illustrate the discretization of the continuity equations. The scheme for Eq. (\ref{eqpc}) is similar.

Let $I_{j}=(z_{j-\frac{1}{2}}$,~$z_{j+\frac{1}{2}})$,~$j=0,1,2,...N$, be a partition of the
computational domain,~$\triangle z_j=z_{j+\frac{1}{2}}-z_{j-\frac{1}{2}}$ and $z_j = \frac{1}{2}(z_{j+\frac{1}{2}}+z_{j-\frac{1}{2}})$. The
{\color{black}finite dimensional subspace} is
$$V=V^h=\bigg\{\phi:\phi|_{I_j}\in \mathbb{P}^k(I_j)\bigg\},$$
where $\mathbb{P}^k$ denotes the polynomials of degree up to $k$ defined on
$I_j$. Both the numerical solution and the test functions will come from
$V^h$.

We use $\mathbb{P}^2$ DG as an example. Choose Legendre polynomials as the basis
functions:
\begin{eqnarray}
  v_0^{(j)}=1,~~v_1^{(j)}=\xi,~~v_2^{(j)}=\frac{3\xi^2-1}{2},
\end{eqnarray}
where $\xi = 2\frac{z-z_j}{\triangle z_j} \in [-1,1]$. The numerical solution can
be written as:
\begin{equation}
n^h(z,t)=\sum_{i=0}^{2}n_j^{(i)}v_{i}^{(j)}, \mbox{~~~~for } z\in I_j.
\end{equation}
It is worth mentioning that the solutions are allowed to
have jumps at the interface $z_{j+\frac{1}{2}}$, and the cell size $\triangle
z_j$ and degree $k$ can {\color{black}vary} from element to element in real applications. These properties
lead to easy adaptivity.

We first introduce three auxiliary variables, {\color{black}$f=n_e v_{e,z}$,} $q=\frac{\partial n_e}{\partial z}$ and $h=\alpha n_e |v_{e,z}|$, then Eq. (\ref{eq1}) reads:
\begin{eqnarray}
  &\frac{\partial n_e}{\partial t}+\frac{\partial (f-D q)}{\partial z} =h,  \label{eq2.10}\\
  & q = \frac{\partial n_e}{\partial z}. \label{eq2.11}
\end{eqnarray}
Multiply Eq. (\ref{eq2.10}) and (\ref{eq2.11}) by a test function $\phi \in V^h$, and
integrate by parts over $I_j$, one gets
\begin{eqnarray}
\int_{I_j}\frac{\partial n_e}{\partial t}\phi \mbox{d}z-\int_{I_j}(f-D q)\phi \mbox{d}z +(f_{j+\frac{1}{2}}-Dq_{j+\frac{1}{2}})\phi_{j+\frac{1}{2}}^{-}\nonumber \\-(f_{j-\frac{1}{2}}-Dq_{j-\frac{1}{2}})\phi_{j-\frac{1}{2}}^{+} =\int_{I_j}h\phi \mbox{d}z, \label{eq12}\\
\int_{I_j}q \phi \mbox{d}z =
n_{e,{j+\frac{1}{2}}}\phi_{j+\frac{1}{2}}^{-}-n_{e,{j-\frac{1}{2}}}\phi_{j-\frac{1}{2}}^{+}-\int_{I_j}n_e\frac{\partial \phi}{\partial z}\mbox{d}z, \label{eq13}
\end{eqnarray}
where $^-$ and $^+$ means the left and right side values at the
interface, respectively. After replacing $n_e$ and $q$ by $n_e^h$ and $q^h$, and choosing
suitable numerical fluxes at the interfaces $z_{j+\frac{1}{2}}$ and $z_{j-\frac{1}{2}}$, Eq.
(\ref{eq12}) and (\ref{eq13}) read:
\begin{eqnarray}
\int_{I_j}\frac{\partial n_e^h}{\partial t}\phi \mbox{d}z-\int_{I_j}(f(n_e^h)-Dq^h)\phi \mbox{d}z+(\widehat{f}_{j+\frac{1}{2}}-D\widehat{q}_{j+\frac{1}{2}})\phi_{j+\frac{1}{2}}^{-}\nonumber\\-(\widehat{f}_{j-\frac{1}{2}}-D\widehat{q}_{j-\frac{1}{2}})\phi_{j-\frac{1}{2}}^{+}=\int_{I_j}h(n_e^h)\phi \mbox{d}z, \\
\int_{I_j}q^h \phi dx
=\widehat{n}_{j+\frac{1}{2}}\phi_{j+\frac{1}{2}}^{-}-\widehat{n}_{j-\frac{1}{2}}\phi_{j-\frac{1}{2}}^{+}-\int_{I_j}n_e^h\frac{\partial \phi}{\partial z}\mbox{d}z.
\end{eqnarray}
where $\widehat{f}_{j\pm\frac{1}{2}}$, $\widehat{q}_{j\pm\frac{1}{2}}$ and $\widehat{n}_{j\pm\frac{1}{2}}$ are the numerical fluxes.

The numerical fluxes are chosen as follows. $\widehat{f}$ is chosen as
the upwind flux since the sign of $v_{e,z}$ is
easy to be obtained, i.e.,
\begin{equation}
\widehat {f}_{j+\frac{1}{2}} =\Bigg\{
\begin{array}{l l}
f_{j+\frac{1}{2}}^-,& \frac{\partial f}{\partial n_e}\geq 0;\\
f_{j+\frac{1}{2}}^+,& \frac{\partial f}{\partial n_e}< 0.
\end{array}
\end{equation}
Of course, other numerical fluxes such as the Lax-Friedrichs flux also works, \begin{equation}
    \widehat {f}_{j+\frac{1}{2}} = \frac{1}{2}( f_{j+\frac{1}{2}}^- + f_{j+\frac{1}{2}}^+) + \max\left|\frac{\partial f}{\partial n_e}\right|(n_e^{h-} - n_e^{h+}).
\end{equation}$\widehat{q}$ and $\widehat{n}$ are chosen according to the alternating
principle:
\begin{equation}
\widehat{q}_{j+\frac{1}{2}} = q^-_{j+\frac{1}{2}},~ \widehat{n}_{j+\frac{1}{2}}=n^+_{e,{j+\frac{1}{2}}} \mbox{~or~} \widehat{q}_{j+\frac{1}{2}} = q^+_{j+\frac{1}{2}},~ \widehat{n}_{j+\frac{1}{2}}=n^-_{e,{j+\frac{1}{2}}}.
\end{equation}

\subsection{The limiter}
Eq. (\ref{eq1}) and Eq. (\ref{eqpc}) are convection dominated, and the electron and positive ion density profiles have steep gradients at the
front of a streamer. Linear high order schemes are quite possible to generate numerical oscillations for this type of problems.
Thus a limiter is greatly desired to eliminate the possible numerical
oscillations.

Chi-Wang Shu proposed a total-variation-bounded minmod limiter which
keeps high order accuracy both in smooth and discontinuous
area \cite{shulimiter}. However, this limiter needs a pre-estimated constant $M$ which is
difficult to be estimated in our problem. Biswas proposed a moment
limiter free of such an constant\cite{biswas}. The limiter is
applied adaptively: first limit the highest-order coefficient, then limit the lower ones
until either a coefficient is found not need to be limited or all
coefficients are limited. The limiter is numerically proved to be able to retain high order
accuracy. Motivated by the moment limiter, Krivodonova proposed a more flexible one, which is derived from the special structure of Legendre polynomials \cite{lilia}.

Write the numerical solution in the form of Legendre polynomials and assume that the numerical solution in cell $k$ is
\begin{equation}
U_k = \sum_{i=0}^p c_i^k P_i(\xi)\label{1dlimiter},
\end{equation}
where $P_i$ is the $i$-th order Legendre polynomials.

Similar to \cite{biswas}, the limiter works from the highest coefficient of the Eq. (\ref{1dlimiter}), replacing $c_i^k$ by
\begin{equation}
\widehat{c_i^k} = \mbox{minmod}( c_i^k, D_i^{+k}, D_i^{-k}) \label{lilialimiter}
\end{equation}
where {\color{black}$D_i^{+k}$ and $D_i^{-k}$ are different approximations to $c_i^k$, and their explicit forms will be given below}, and {\color{black} the minmod function} is defined as follows:
\begin{equation}
\mbox{minmod}(a,b,c) = \left\{
\begin{array}{l l}
\min(a,b,c),& a\geq 0, b\geq 0, c\geq 0;\\
\max(a,b,c),& a< 0, b< 0, c< 0;\\
0,& \mbox{otherwise}.
\end{array}\right.
\end{equation}

To complete the limiter Eq. (\ref{lilialimiter}), it only leaves to find proper definitions for $D_i^{+k}$ and $D_i^{-k}$. This can be achived by forward or backward finite differences.

For Eq. (\ref{1dlimiter}), taking its ($i$-1)-th and $i$-th derivatives, one gets
\begin{eqnarray}
&\frac{\partial^{i-1}U_k}{\partial x^{i-1}}= (\frac{2}{\triangle x})^{i-1}(2i-3)!!c_{i-1}^k+(\frac{2}{\triangle x})^{i-1} \frac{\partial^{i-1}}{\partial \xi^{i-1}} \sum_{m>i-1}^p c_m^k P_m(\xi), \\
&\frac{\partial^{i}U_k}{\partial x^{i}}= (\frac{2}{\triangle x})^{i}(2i-1)!!c_{i}^k+(\frac{2}{\triangle x})^{i} \frac{\partial^{i}}{\partial \xi^{i}} \sum_{m>i}^p c_m^k P_m(\xi).  \label{dir}
\end{eqnarray}
By finite difference, one gets
\begin{eqnarray}
\frac{\frac{\partial^{i-1}U_{k+1}}{\partial x^{i-1}}-\frac{\partial^{i-1}U_k}{\partial x^{i-1}}}{\triangle x} &=& (\frac{2}{\triangle x})^{i}\frac{(2i-3)!!}{2}(c_{i-1}^{k+1}-c_{i-1}^{k})+ (\frac{2}{\triangle x})^{i}  \nonumber \\ &&\frac{\partial^{i-1}}{\partial \xi^{i-1}} \sum_{m>i-1}^p \frac{1}{2}(c_m^{k+1}-c_m^k)P_m(\xi). \label{dif}
\end{eqnarray}
Considering Eq. (\ref{dir}) and Eq. (\ref{dif}) together, and noticing they are the exact or approximate form of $\frac{\partial^{i}U_k}{\partial x^{i}}$, one can get
\begin{equation}
c_i^k = \frac{c_{i-1}^{k+1}-c_{i-1}^k}{2(2i-1)}+O(\triangle x^{i+1}). \label{ci1}
\end{equation}
Similarly,
\begin{equation}
c_i^k = \frac{c_{i-1}^{k}-c_{i-1}^{k-1}}{2(2i-1)}+O(\triangle x^{i+1}). \label{ci2}
\end{equation}
By Eq. (\ref{ci1}) and Eq. (\ref{ci2}), we can define
\begin{eqnarray}
D_i^{-k}=\frac{c_{i-1}^k - c_{i-1}^{k-1}}{2(2i-1)}, ~~D_i^{+k}=\frac{c_{i}^{k+1} - c_{i-1}^{k}}{2(2i-1)}.
\end{eqnarray}
The above definitions may lead to too much numerical diffusions, thus a variable $\alpha_i$ is introduced to adjust the numerical diffusions, i.e.,
\begin{equation}
\widehat{c_i^k}=\mbox{minmod}( c_i^k, \alpha_i (c_{i-1}^{k+1}-c_{i-1}^k), \alpha_i( c_{i-1}^k-c_{i-1}^{k-1})).
\end{equation}
where $\alpha_i$ is a variable satisfing
\begin{equation}
\frac{1}{2(2i-1)} \leq \alpha_i \leq 1.
\end{equation}
In actual applications, we may use $\alpha_i=1$.

Start from the highest order coefficient, i.e., $p$-th order, replace $c_i^k$ with $\widehat{c_i^k}$, stop until finding the first coefficient that need not be replaced or the the limiting process reaches the lowest order coefficient. In this way, the limiting process completes.

\subsection{Temporal discretization of the continuity equation}
The auxiliary variable $q$ can be locally solved from Eq. (\ref{eq13})
from element to element. Plugging $q$ into Eq. (\ref{eq12}), results
in an ODE:
\begin{equation}
\frac{\mbox{d}n^h}{\mbox{d}t}=\mbox{L}(n^h)
\end{equation}

The ODE can be solved by any time marching scheme. We
choose the third order total-variation-diminishing Runge-Kutta
scheme~\cite{ShuTVD}:
\begin{eqnarray}
n^{(1)}&=&n^n+\mbox{L}(n^n)\triangle t, \\
n^{(2)}&=&\frac{3}{4}n^{n}+\frac{1}{4}(n^{(1)}+\mbox{L}(n^{(1)})\triangle t), \\
n^{(3)}&=&\frac{1}{3}n^{n}+\frac{2}{3}(n^{(2)}+\mbox{L}(n^{(2)})\triangle t), \\
n^{n+1}&=&n^{(3)}.
\end{eqnarray}
The limiter should be applied at each sub-step of the Runge-Kutta scheme.

\section{Validation and comparisons}
We list several examples to validate the method and the codes, and make some comparisons with other methods to show the high
resolution and high order properties of DG for streamer simulations.
\subsection{A smooth example of convection-diffusion equations}\label{sina}
The following problem was tested with a sufficient small time step and
$\mathbb{P}^2$ DG:
\begin{eqnarray}
\frac{\partial n}{\partial t}+\frac{\partial n}{\partial x}
-\frac{\partial ^2n}{\partial x^2}=0,
~~n(x,0)=\sin(x),~~ x\in [0,2\pi],
\end{eqnarray}
with periodic boundary condition. The exact solution is
$$n(x,t)=e^{-t}\sin(x-t),\qquad x\in[0,2\pi].$$
The results listed in Table 1 show that the
numerical solution agrees well with the exact solution and optimal
convergence orders for both $n$ and its derivative $q=\frac{\partial n}{\partial x}$ are
achieved.

\begin{table}[ht] \label{tab51}
\caption{Error and convergence analysis for example \ref{sina}}
\centering
\begin{tabular}{c c c c c}
\hline
mesh size &  max$|n-n^h|$ & order&  max$|q-q^h|$&   order \\
\hline
$2\pi/20$ & 8.95354e-5 & - & 2.15846e-4 & - \\
$2\pi/40$ & 1.15799e-5 & 2.9508 & 2.68188e-5 & 3.0087 \\
$2\pi/80$ & 1.47143e-6 & 2.9763 & 3.33401e-6 & 3.0079 \\
$2\pi/160$& 1.85488e-7 & 2.9882 & 4.05052e-7 & 3.0411 \\
\hline
\end{tabular}
\end{table}

\subsection{An advection problem with discontinuities}\label{shock}
The following convection problem, consisting of a Gaussian curve, a
unit square impulse, a triangle and a semi-ellipse, was used to test
the {\color{black} ability of capturing discontinuity}~\cite{moving mesh}:
\begin{eqnarray}
 && \frac{\partial n}{\partial t}+\frac{\partial n}{\partial x}=0,\qquad
  x\in[-1,+1], \\
&&n(x,0)=\left \{ \begin{array}{l l}
\exp\left(\frac{-\ln 2}{36 \times 0.005^2}(x+0.7)^2\right),&-0.8 \leq x \leq -0.6,\\
1,&-0.4 \leq x \leq -0.2,\\
1-10|x-0.1|, & 0 \leq x \leq 0.2,\\
\sqrt{1-10^2(x-0.5)^2}, & 0.4 \leq x \leq 0.6,\\
0, & \mbox{otherwise};
\end{array} \nonumber
\right.
\end{eqnarray}
with periodic boundary condition.

\begin{figure}[!h]
\centering
\subfigure[results with 100 elements.]{ \label{fig:5.2:a} 
\includegraphics[width= 0.485 \textwidth]{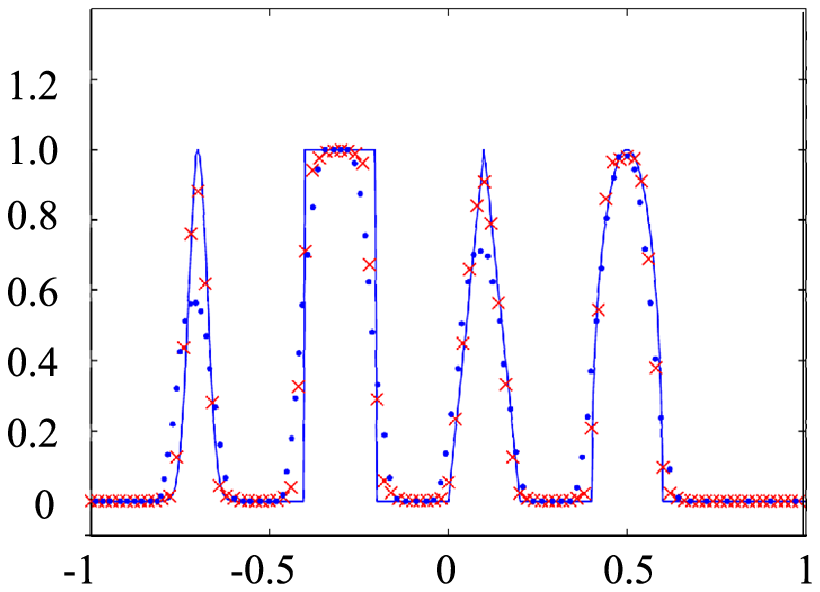}}
\subfigure[results with 200 elements.]{
\label{fig:5.2:b} 
\includegraphics[width= 0.485 \textwidth]{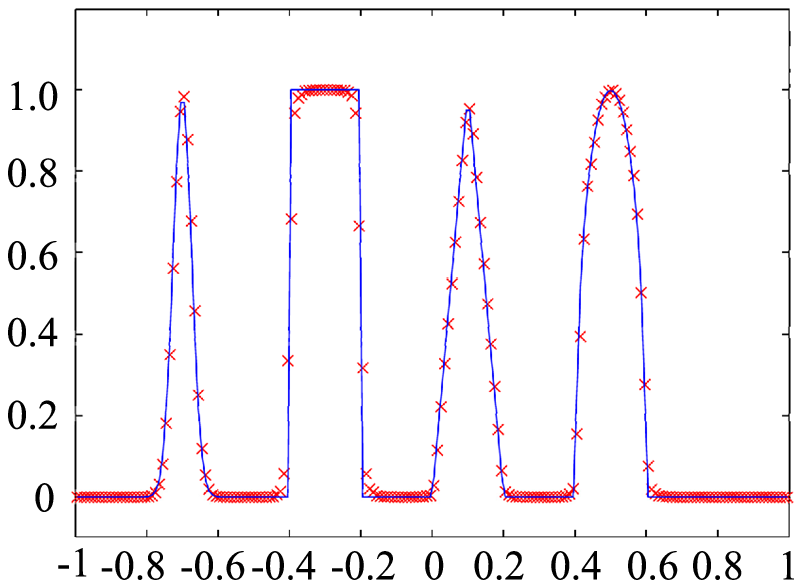}}
\caption{Comparison of different methods for example \ref{shock} with discontinuities. (solid: exact solution; dot: FVM with moving mesh method; cross: discontinuous Galerkin method)}
\label{fig:5.2}
\end{figure}

Results are shown in Figure \ref{fig:5.2}. Compared with third order FVM combined with moving-mesh, third order DG gets a
better solution using the same number of mesh points (cf.  Figure \ref{fig:5.2:a}).
For the Gaussian curve, DG and moving-mesh FVM get the extreme maximum of 0.88 and less than 0.6, respectively, while the exact
value is 1.0. Those for the triangle are 0.91 and less than 0.7,
respectively, while the exact value is 1.0. Results for the curve of
the unit square impulse and semi-ellipse are comparable.

If 200 mesh points are used in DG, the numerical solution of DG would be further improved (cf.
Figure \ref{fig:5.2:b}).

\subsection{Davies' test}\label{secdavies}
\begin{figure}[!h]
\centering
\subfigure[the numerical solution by DG at $t=0.4T$]{ \label{fig:subfig:a} 
\includegraphics[width= 0.485 \textwidth]{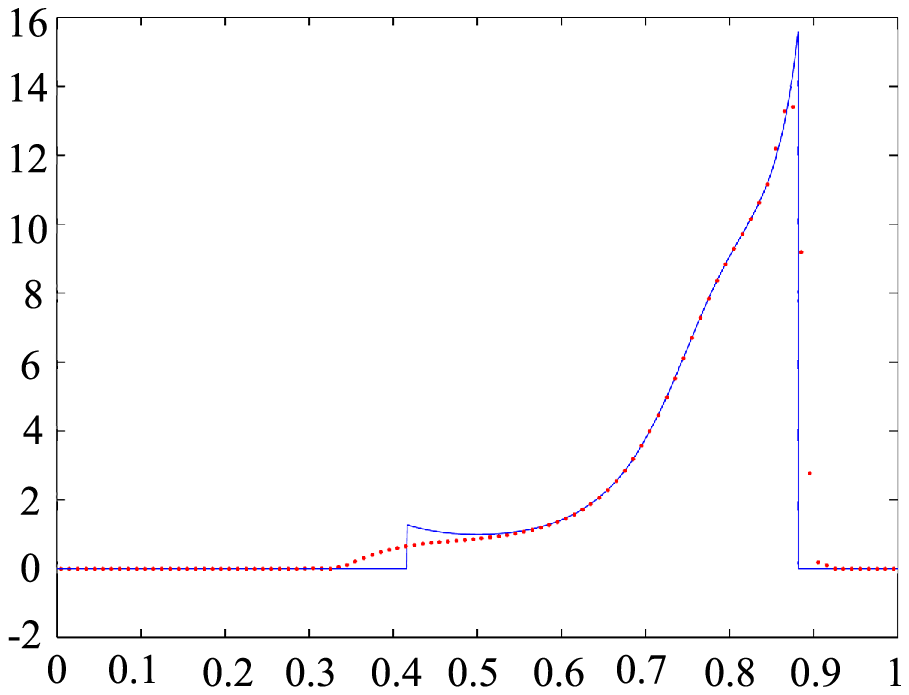}}
\subfigure[the numerical solution by DG at $t=T$]{
\label{fig:subfig:b} 
\includegraphics[width= 0.485 \textwidth]{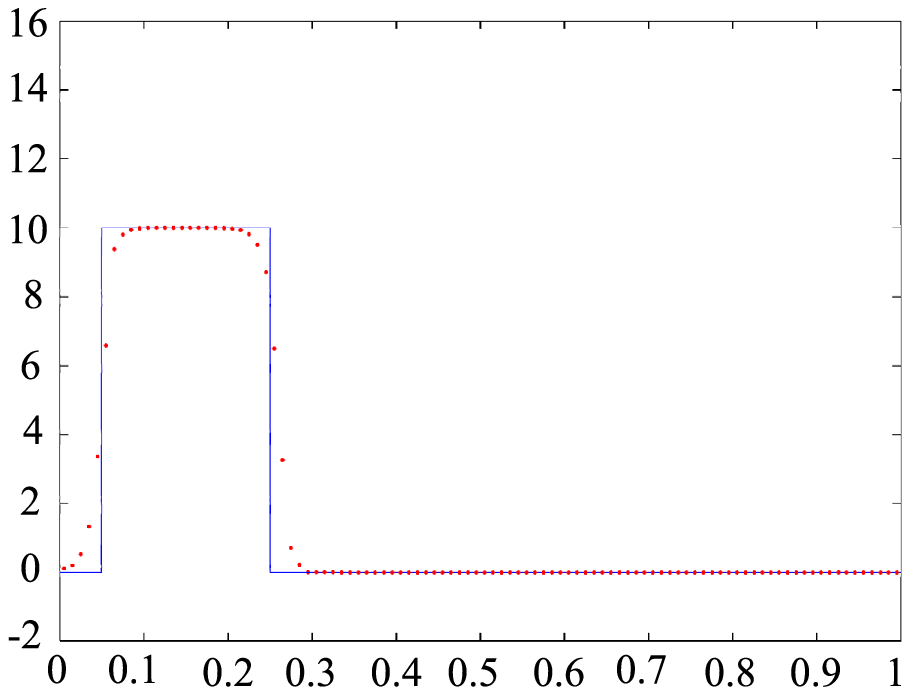}}
\caption{Comparison of different methods for example \ref{secdavies} with discontinuity. (solid: exact solution; dot: discontinuous Galerkin method)}
\label{fig:davies}
\end{figure}

Davies' test was used to test the ability of capturing the discontinuity\cite{criticalanalysis}. Its exact solution is similar to the profile of the charge density in a streamer channel, and the convection speed of the wave is also a function of position, which is similar to the case of streamer discharge simulations:
\begin{eqnarray} \label{daviestest}
\frac{\partial n}{\partial t}+\frac{\partial(nv_z)}{\partial z}=0, v_z = 1+9\sin^8(\pi z),z\in [0,1],  \\
n(z,0)= \left \{ \begin{array}{l l}
        10,& 0.05 \leq z \leq 0.25\\
        0, & \mbox{otherwise}.
         \end{array}
\right.
\end{eqnarray}
The period time of the wave is $T\approx 0.591$. The numerical solution of DG under the same configuration as \cite{criticalanalysis} is shown in Figure \ref{fig:davies} and the numerical solutions obtained by FVM-MUSCL and FEM-FCT were shown in Figure 1(b) of \cite{criticalanalysis}.

At $t=0.4T$, DG captures the discontinuity within 5 cells, and the computed maximum is 14, while the maximum obtained by FVM-MUSCL and FEM-FCT are around 12. The exact is around 16. In addition, the solution obtained by DG is free of oscillations while those of FVM-MUSCL and FEM-FCT have small oscillations.

At $t=T$, the average errors of FVM-MUSCL and FEM-FCT are 0.2650 and 0.2677 \cite{criticalanalysis}, respectively, while that of DG is 0.2272. In addition, FVM-MUSCL and FEM-FCT have small overshoots while DG doesn't.

\section{Simulation results}

A parallel-plate double-headed streamer discharge in nitrogen using 1.5-dimensional fluid model was simulated.
The photo-ionization was considered by a background ionization in the initial condition.

The two plates are paralleled, and are perpendicular to the $z$-axis. On the anode ($z=0$), 52 kV voltage was applied, and the cathode ($z=1$ cm) was grounded. The charge was assumed distributing uniformly on discs with a radius of 0.05 cm. The initial condition was as follows:
\begin{equation}
n_e(z,0) = n_b +n_0 \exp(-(\frac{z-z_0}{\sigma_z})^2),
\end{equation}
where $z_0 = 0.5$ cm, $n_0 = 10^{14}$ cm$^{-3}$, $\sigma_z = 0.027$ cm, and the background pre-ionization $n_b = 10^8$ cm$^{-3}$.

\begin{figure}[!h]
\centering
\includegraphics[width=0.6\textwidth]{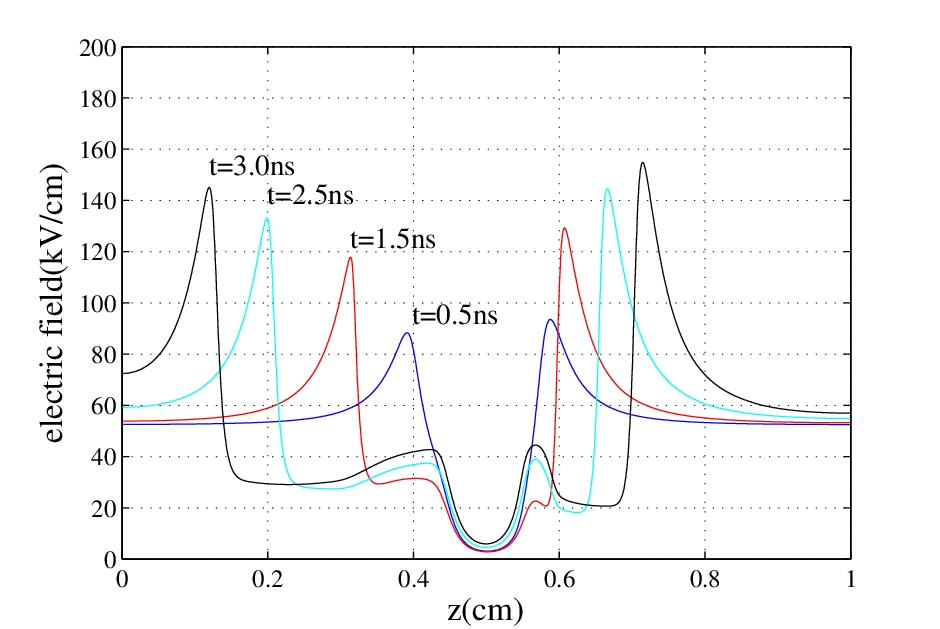} \\
  \caption{the electric field evolution of a double-headed streamer.}
  \label{ef}
\end{figure}

After the voltage applies, the cathode-directed (positive) and anode-directed (negative) (the right half and the left half in Figure \ref{ef}, respectively) develop to the opposite electrodes immediately. At $t=2.5$ ns, the negative and positive streamer moves 0.28 cm and 0.18 cm, respectively.
In fact, the propagation speed of streamers varies with time. When $t\leq 3$ ns, the velocities of the negative and positive streamer are around
$~0.8-1.8\times 10^8$ cm/s and $0.4-1.0\times 10^8$ cm/s, respectively.

\begin{figure}[!h]
\centering
\includegraphics[width=0.6\textwidth]{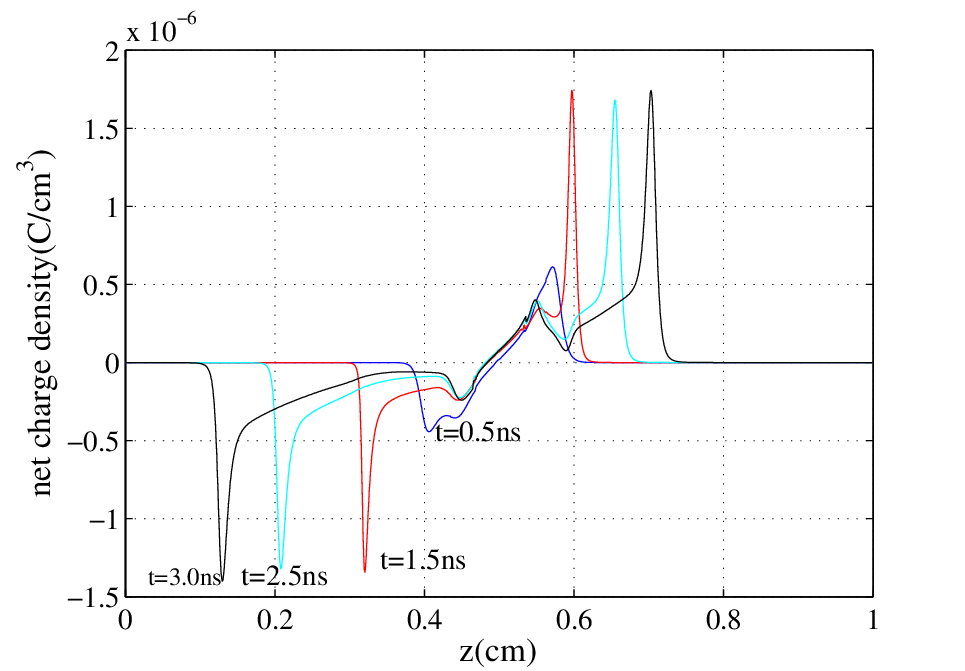} \\
  \caption{the net charge densities evolution of a double-headed streamer.}
  \label{netcharge}
\end{figure}

As is shown in Figure \ref{netcharge}, at $t=2.5$ ns, the maximums of the net charge densities in the negative and positive streamer channels are $1.30$ and $1.65$ $\mu$C/cm$^3$, respectively. At the front of each streamer, there is a $0.1-0.2$ mm thick layer with a much larger net charge density. This conclusion coincides with previous work.

\section{Conclusions}
The 1.5-dimensional streamer discharge model retains the basic intergradients of a discharge process.
In this paper, a local discontinuous Galerkin method for the 1.5-dimensional streamer discharge simulations is proposed. The electric field in the discharge channel is solved analytically and the continuity equations are solved by the local discontinuous Galerkin method with a limiter.
A 1-cm parallel-plate double-headed streamer is simulated. The preliminary results suggest the potential of the local discontinuous Galerkin method for streamer discharge simulations.

A discharge model considering a detailed photo-ionization process is under working and the 2-dimensional simulations using the local discontinuous Galerkin method will be reported later.

\section*{Appendix}
\begin{figure}[!h]
\centering
\subfigure[ numerical solution by first order upwind scheme.]{ \label{fig:app:a} 
\includegraphics[width= 0.485 \textwidth]{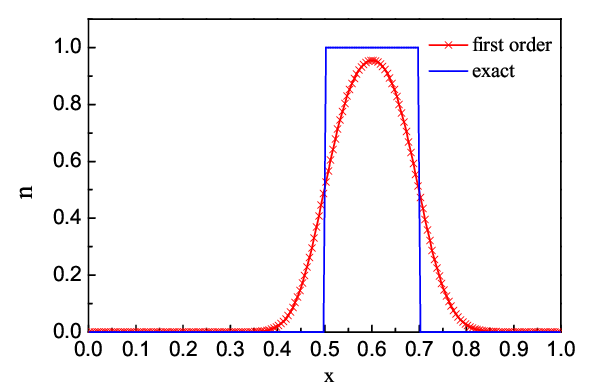}}
\subfigure[numerical solution by second order central finite difference scheme.]{
\label{fig:app:b} 
\includegraphics[width= 0.485 \textwidth]{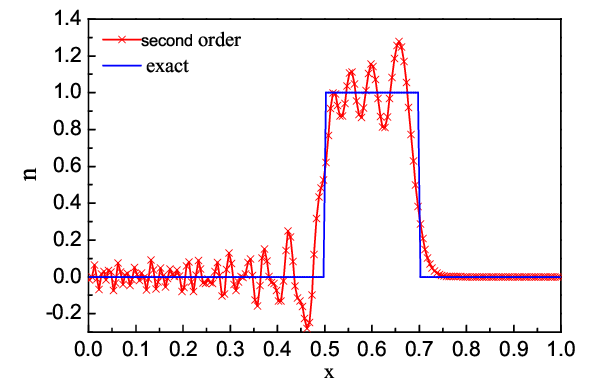}}
\caption{the numerical diffusion and oscillation of linear schemes}
\label{fig:app}
\end{figure}

Consider a simple advection problem $\frac{\partial n}{\partial t}+\frac{\partial n}{\partial x}=0$ with periodic boundary conditions. If the initial condition is a square, then the exact solution is a square wave moving from the left to the right. The numerical solutions obtained by the first order upwind finite difference scheme and second order central difference scheme are shown in Figure \ref{fig:app}, which shows first order scheme has numerical diffusions and a high order ($\geq 2$) linear scheme has numerical oscillations.

\section*{Acknowledgement}
This work is supported by National Basic Research Program of China (973 program)(No.
2011CB209403) and National Natural Science Foundation of China (No. 51207078).

We thank the reviewer for carefully reading the manuscript which leads to a great improvement of this paper.



\bibliographystyle{elsarticle-num}







\end{document}